\documentclass[floatfix,preprint,authoryear]{elsarticle}
\usepackage{amsmath, amsthm, amssymb}
\usepackage{graphicx}

\journal{Journal of Theoretical Biology}

\begin{document}

\def\g{\gamma}
\def\r{\rho}
\def\w{\omega}
\def\wo{\w_0}
\def\wp{\w_+}
\def\wm{\w_-}
\def\t{\tau}
\def\av#1{\langle#1\rangle}
\def\pf{P_{\rm F}}
\def\pr{P_{\rm R}}
\def\F#1{{\cal F}\left[#1\right]}

\begin{frontmatter}

\title{The effect of habitats and fitness on species coexistence in systems with cyclic dominance}

\author[vtais]{Ryan Baker}

\author[vtp,vtcsmbp,vtais]{Michel Pleimling\corref{cor1}}
\ead{pleim@vt.edu}

\cortext[cor1]{Corresponding author}

\address[vtais]{Academy of Integrated Science, Virginia Polytechnic Institute and State University, Blacksburg, Virginia 24061-0405, USA}
\address[vtp]{Department of Physics, Virginia Polytechnic Institute and State University, Blacksburg, Virginia 24061-0435, USA}
\address[vtcsmbp]{Center for Soft Matter and Biological Physics, Virginia Polytechnic Institute and State University, Blacksburg, Virginia 24061-0435, USA}

\date{\today}

\begin{abstract}
Cyclic dominance between species may yield spiral waves that are known to provide a mechanism enabling persistent species coexistence. 
This observation holds true even in presence of spatial heterogeneity in the form of quenched disorder. In this work we study the effects
on spatio-temporal patterns and species coexistence of structured spatial heterogeneity in the form of habitats that locally provide one
of the species with an advantage. Performing extensive numerical simulations of systems with three and six species we show that these structured habitats
destabilize spiral waves. Analyzing extinction events, we find that species extinction probabilities display 
a succession of maxima as function of time, that indicate a periodically enhanced probability for species extinction. Analysis of the mean
extinction time reveals that as a function of the parameter governing the advantage of one of the species a transition between stable
coexistence and unstable coexistence takes place. We also investigate how efficiency as a predator or a prey affects species coexistence.
\end{abstract}


\end{frontmatter}


\section{Introduction}
Understanding the emergent spatial structure in ecological networks is important in order to assess the system's
stability and resilience against species extinction \citep{May74,Smith74,Smith82,Hofbauer98,Nowak06,Dobramysl18}.
A crucial role is played by the interaction network, and many recent studies have focused on space-time patterns
and their importance for species coexistence when considering multiple species with complex relationships, see
\citep{Frey10,Szolnoki14,Dobramysl18} for some recent reviews. Already three mobile species in cyclic competition spontaneously
form spiral waves \citep{May75} which results in an enhanced resilience against species extinction.
This enhanced stability can be jeopardized by a high mobility that tends to break up the space-time patterns \citep{Kerr02,Kirkup04,
Reichenbach07a,Reichenbach07b,Reichenbach08a}. Even richer patterns emerge for larger numbers of interacting 
species as well as for more complex interaction schemes \citep{Szabo01a,Szabo01b,Szabo04,Szabo05,Szabo07a,Szabo07b,Szabo07c,
Perc07,Szabo08a,Szabo08b,Roman12,Avelino12a,Avelino12b,Roman13,Vukov13,Mowlaei14,Cheng14,Avelino14a,Avelino14b,Szolnoki15,Roman16,
Labavic16,Brown17,Avelino17,Esmaeili18,Avelino18,Szolnoki18,Danku18,Brown19,Avelino19}.

Many of these studies of emerging space-time patterns in ecological networks considered the simplest possible setup with 
species-independent rates that are homogeneous in space and that do not evolve over time, but instead remain the same generation after 
generation. Of course, this type of simplification is important in order to elucidate generic properties of pattern formation,
biodiversity, and species extinction. Still, as many relevant aspects of real-world ecologies (heterogeneity of the
physical environment, species fitness, seasonal changes, evolutionary adaptation, etc.) are being ignored in a standard setup, it is a valid question to ask  
whether ecological stability and species extinction are affected when increasing the realism of the system.

In this work we focus on two systems, the three-species May-Leonard model \citep{May75} as well as a six-species model \citep{Roman13},
that both display in a standard setup the spontaneous formation of spiral patterns. These spirals emerge because of
a cyclic interaction scheme that results 
in the formation of spiral arms where a front dominated by one
species is followed by a front dominated by the only species that is not a prey of the first species. As shown in earlier studies, 
propagating spiral waves enhance the system's stability and foster species coexistence \citep{Reichenbach07a,Roman13}. 
In variance with the standard setup, we consider in the following a heterogeneous environment in the form of a habitat structure. Inside
a habitat one of the species has an advantage over the others as individuals from this species have a higher
probability to survive an attack. We also consider the situation where individuals are characterized by the efficiency of their
predation and escape capabilities. In our implementation of evolutionary adaptation the
efficiency of the parent is inherited by the off-spring, with small random changes that
reflect differentiation and adaptation due to mutations. Our goal is to develop an understanding of species extinction
in systems with stabilizing spiral waves subjected to spatial (not random, but structured) heterogeneity and temporal evolution of species fitness.

Our paper is organized in the following way. In Section 2 we provide a detailed discussion of our modifications to the 
standard three- and six-species models with emerging spiral waves. Our main focus has been on the three-species case, and
Section 3 provides a detailed overview of the results we obtained for this case. In Section 4 we verify through a discussion
of the six-species system that our main conclusions from Section 3 are also valid for more complicated situations. We conclude in Section 5.

\section{Model}
\begin{figure}
 \centering \includegraphics[width=0.6\columnwidth,clip=true]{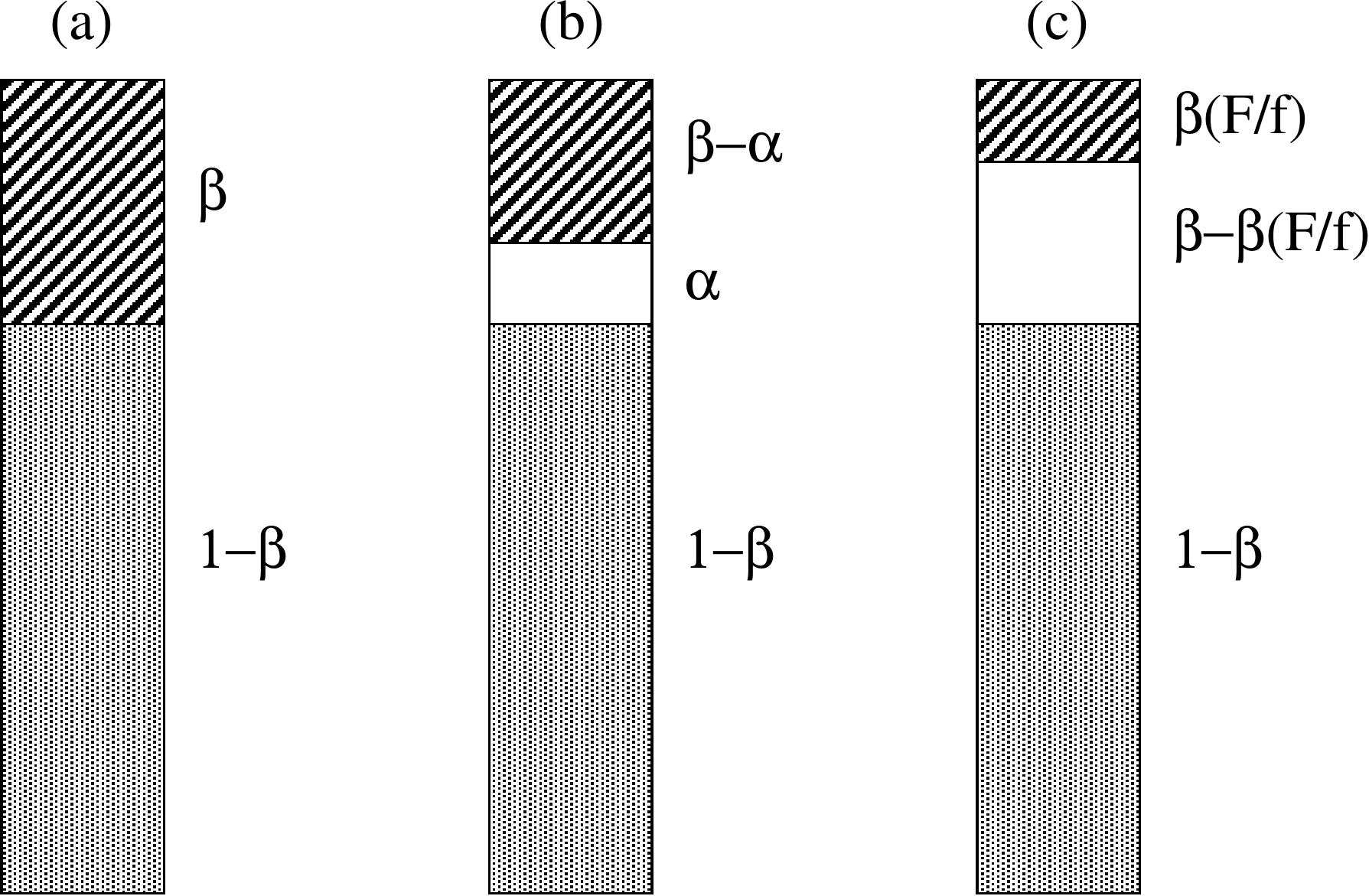}
\caption{(a) The standard scheme where a predator-prey pair interaction results in either the removal of the prey with probability $\beta$
(striped area) or the exchange of positions with probability $1 - \beta$ (gray area). (b) In its habitat an individual from species 1 has a probability $\alpha$ 
(with $0 < \alpha \le \beta$) to remain
unharmed when attacked by a predator (white area: no update to the system) or a probability $\beta - \alpha$ for being removed. (c) The efficiency $f$
respectively $F$ provided to the prey respectively predator results for the case $F < f$ in a non-zero probability $\beta - \beta \left( F/f \right)$ 
for the prey to survive the attack of the predator and to remain on the same lattice site.
}
\label{fig1}
\end{figure}

We consider as our starting point a broad family of May-Leonard type predator-prey models
with cyclic interactions. Using notation established in previous work \citep{Roman13,Mowlaei14,Roman16,Labavic16,Brown17,Esmaeili18},
we call $(N,r)$ the game consisting of $N$ species where each species attacks $r$ others in a cyclic manner. 
In its simplest version this game, which is played on a 
two-dimensional lattice, consists in the following reactions, taking place between two neighboring sites (see (a) in Figure \ref{fig1}):
\begin{eqnarray}
s_{i} + s_{j} & \xrightarrow[]{\beta} & s_{i} + \emptyset \label{eq:1} \\
s_{i} + \emptyset & \xrightarrow[]{\beta} & s_{i} + s_{i} \label{eq:2}\\
s_{i} + X & \xrightarrow[]{1-\beta} & X+s_{i} \label{eq:3}
\end{eqnarray}
where $s_i$ is an individual of species $i$, whereas $\emptyset$ indicates an empty site and $X$ can be an empty site or a site occupied by
an individual from any species. This reaction scheme encompasses predation events (\ref{eq:1}), where the prey $s_j$ on a site neighboring the predator 
$s_i$ is removed, birth of off-springs on a neighboring empty site (\ref{eq:2}), as well as mobile individuals, see equation (\ref{eq:3}),
that either can jump to a neighboring empty site or swap places with one of their neighbors.

The $(N,r)$ reaction scheme yields a variety of situations, depending on the values of $N$ and $r$. In this work we are interested in situations
that lead to the spontaneous formation of spirals where each propagating wave front is dominated by individuals of one species. The two
cases we will discuss in detail in the following are the (3,1) case, which is identical to the celebrated May-Leonard model \citep{May75,Frey10,Dobramysl18},
and the (6,4) case \citep{Roman13}, a game where six species interact in such a way that for each species there is exactly one other species
that is not its prey. In a two-dimensional lattice both interaction schemes result in the
spontaneous formation of spiral waves. Each spiral arm is dominated by one species, with the
following spiral arm containing mainly individuals of the species that is not prey of the first one.

In many studies it is assumed that reaction rates are homogeneous in space and time and also the same for every species. While it is important to
consider the simplest version of a model for gaining an understanding of its properties, a more refined approach is needed in order to increase
the realism of the model. We consider in the following two modifications of the standard $(N,r)$ model on a lattice, 
as we add a habitat structure that gives one of the
species an advantage in escaping predation, and provide each individual at its birth with a fitness (efficiency) that changes the probability that the prey
survives the attack of a predator. We do not change the rate associated with the mobility and the birth rate, keeping them for every 
individual and at all times at fixed values $\beta$ and $1-\beta$, respectively. The modifications discussed below only affect the rate at which
a prey is removed during a predator-prey interaction. In this way we not only minimize the number of free parameters in our model, we also make sure 
that we only compare situations with a very similar width of the spiral structure, as this width depends strongly on the mobility of the
individuals \citep{Reichenbach07a}.

We divide our lattice in 16 even squares where 8 of them, arranged in a checkerboard, are the habitats of species 1 and provide this species
with an advantage evading the attacks of its predators. Whereas in the 8 outside areas all species are equal and interact with the same rate $\beta$,
see equations (\ref{eq:1})-(\ref{eq:3}) and Figure \ref{fig1}(a), inside their habitats individuals of species 1 have a reduced probability to be
removed when attacked by a predator. As shown in Figure \ref{fig1}(b), this is realized by changing for a prey belonging to species 1 the probability
for being removed to $\beta - \alpha$ where the habitat modifier $\alpha$ has a fixed value chosen before the beginning
of the simulation from the interval  $\left[0, \beta \right]$. It follows that
for each attack on a individual of species 1 there is the probability $\alpha$ that the individual escapes unharmed, resulting in no changes to the system.
This setup strongly differs from the metapopulation model for the rock-paper-scissors game presented in recent papers \citep{Nagatani18a,Nagatani18b}
as well as from earlier studies of the May-Leonard model with spatially variable random rates \citep{He10,He11}.

Of course, some limiting cases are immediately obtained. For $\alpha =0$ we recover the standard model, whereas for $\alpha = \beta$ species 1 is
perfectly safe in its habitats and no attack on an individual from that species will be successful.

In order to make our particle-based description more realistic, we introduce individual fitness and evolutionary adaptation by
endowing every individual with a number $f$ that reflects the varying
efficiency of their predation and escape capabilities. 
Our setup follows general ideas introduced in \citep{Dobramysl13,Chen18} for predator-prey systems with one or two predators, 
but differs in the details of the implementation.
This efficiency is assigned to an individual at its birth (or at the beginning of
the simulation in case the individual is present when starting the run). Individuals inherit their efficiency from their parent, but we allow for
some random changes that reflect differentiation and adaptation due to mutations. This is realized by generating the efficiency of the off-spring
from a Gaussian distribution with variance 1 centered at the efficiency of the parent. When setting up the system we assign a common efficiency $f_{init}$ to
all individuals present at the start of the simulation. As we do not change the width of the distribution, it is through the value of $f_{init}$ that we 
control the relative change of the efficiency at the birth of an off-spring, i.e. for small values of $f_{init}$ the relative change is large, whereas for 
large values of $f_{init}$ the relative change is small. The case of an efficiency that does not change over time is recovered in the limiting case
$f_{init} \longrightarrow \infty$. 

Let us denote as $f$ respectively $F$ the efficiencies of the prey respectively the predator that have been selected for an interaction. If the
prey is from a species different than species 1 or is a member of species 1 that is not inside one of its habitats,
then the probability that the prey is removed is changed to $\beta \left( F/f \right)$ when $F < f$, i.e. because of the larger efficiency of the prey there is
the non-zero probability $\beta - \beta \left( F/f \right)$ for the prey to survive the attack of the predator and to remain on its lattice
site, see Figure \ref{fig1}(c). For $F > f$ the standard scheme Figure \ref{fig1}(a) prevails where predator and prey are swapping places with probability 
$1- \beta$ and the prey is removed with probability $\beta$. On the other hand, if the prey is a member of species 1 located in its habitat, then
we have the interesting situation that, depending on the value of the ratio $F/f$, either the ability of the prey to escape its attacker is
further increased (for $F/f < 1$) or the efficiency of the predator reduces the odds of the prey to remain unharmed (for $F/f > 1$). Indeed,
in the first case the combination of the habitat advantage and the larger efficiency of the prey increases the prey's probability to survive the
attack unharmed to $\mbox{min}\left[\alpha + \beta - \beta \left( F/f \right), \beta \right]$. In the second case, however, the efficiency of the predator
works against the advantage of the prey of being in its habitat, which reduces the probability of surviving the attack to 
$\mbox{max} \left[ \alpha + \beta - \beta \left( F/f \right), 0 \right]$.

In our agent-based simulations we prepare the system initially in a disordered state where with probability $\frac{1}{N+1}$ every lattice site is 
either occupied by a member of one of the $N$ species or is left empty. We then select randomly a lattice site and one of its four
nearest neighbors. In the standard setup, i.e. without habitats and fitness, if the two selected sites are occupied by individuals 
from two species that have a predator-prey relationship and if the first selected site is occupied by a predator, then with probability $\beta$ 
the prey is removed from the other lattice site, whereas with probability
$1 - \beta$ predator and prey swap places. If on the other hand the species involved have a neutral relationship,
i.e. do not have a predator-prey relationship, or if the first selected site is not occupied by a predator,
then only the swapping takes place with probability $1- \beta$. For the case that exactly one of the selected sites is unoccupied, then the individual 
located on the occupied site jumps with probability $1- \beta$ into the open site, otherwise it deposits an offspring into
this unoccupied lattice site. 
In the modified versions with habitats and/or fitness, we keep $\beta$ constant, but modify the probability of a successful attack
as described above. For all the simulations presented in this work we fixed $\beta = 0.3$

We define as one time step $L \times L$ proposed updates that start with a randomly selected site,
with $L$ being the linear size of the system. The system sizes discussed in the following range from $L=16$ to $L =56$.

The focus on our work is on the time evolution of the system and on extinction events. In order to obtain data that allow us to
discuss probability distributions, for every set of parameters (system size $L$, habitat modifier $\alpha$, initial fitness $f_{init}$)
we perform millions of runs with different random initial conditions and different random number sequences.

\section{The three-species case}
The influence of spatial degrees of freedom and mobility on species evolution and ecosystem self-organization is well
established \citep{Turing52,Koch94,Levin76,Hassel94,Maron97,Weber14}. Cyclic competition between species has been 
discussed as one possible mechanism responsible for persistent species coexistence. In \citep{Kerr02} it was shown that
in a spatial setting (i.e. a Petri dish) three cyclically competing strains of bacteria coexist for a long time,
whereas in the well-mixed environment of shaken flasks two species go extinct quickly. Both theoretical 
\citep{Reichenbach07a,Reichenbach07b,Reichenbach08a,He11,Lamouroux12,Rulands13,Szczesny13,Szczesny14,Szolnoki14,Dobramysl18} and experimental
\citep{Siegert95,Igoshin04} studies have revealed that spatio-temporal patterns, as for example propagating fronts or spiral waves, often
accompany long-lasting coexistence between different species.

As in finite systems fluctuations ultimately yield species extinction, it has been proposed that the dependence of the characteristic extinction time $\tau$
on the system size allows to probe the stability of a system \citep{Antal06,Reichenbach07a,Reichenbach08b}. 
In our systems with $L^2$ sites the following three scenarios are possible: if $\tau /L^2 \longrightarrow \infty$ for $L^2
\longrightarrow \infty$, coexistence is stable and extinction takes very long, whereas if $\tau /L^2 \longrightarrow 0$ for $L^2
\longrightarrow \infty$, coexistence is unstable, extinction is fast, and the probability for species extinction taking place approaches 1.
These two cases are separated by the case of neutral stability for which $\tau \sim L^2$.

\begin{figure}
 \centering \includegraphics[width=0.8\columnwidth,clip=true]{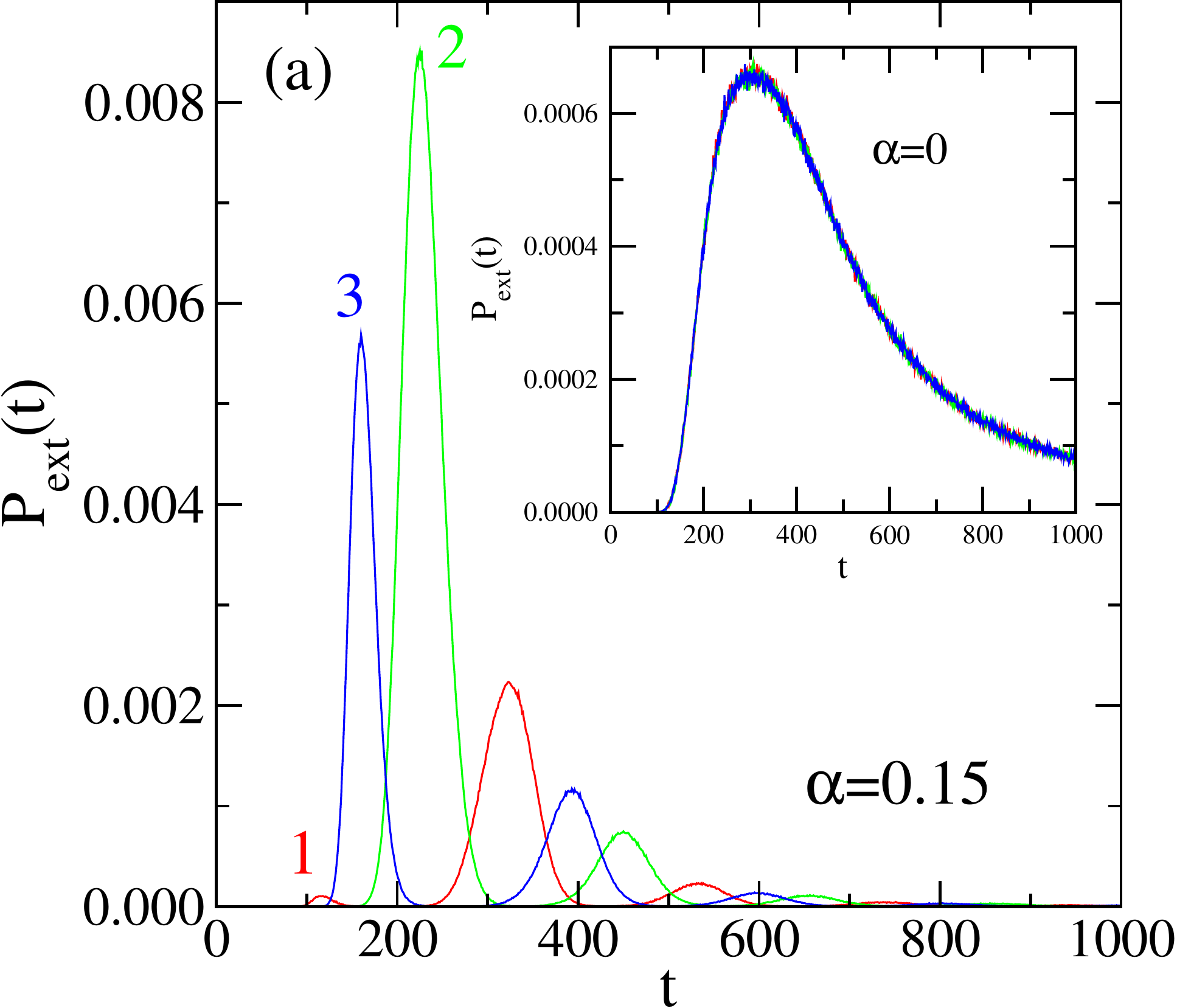}\\
\includegraphics[width=0.77\columnwidth,clip=true]{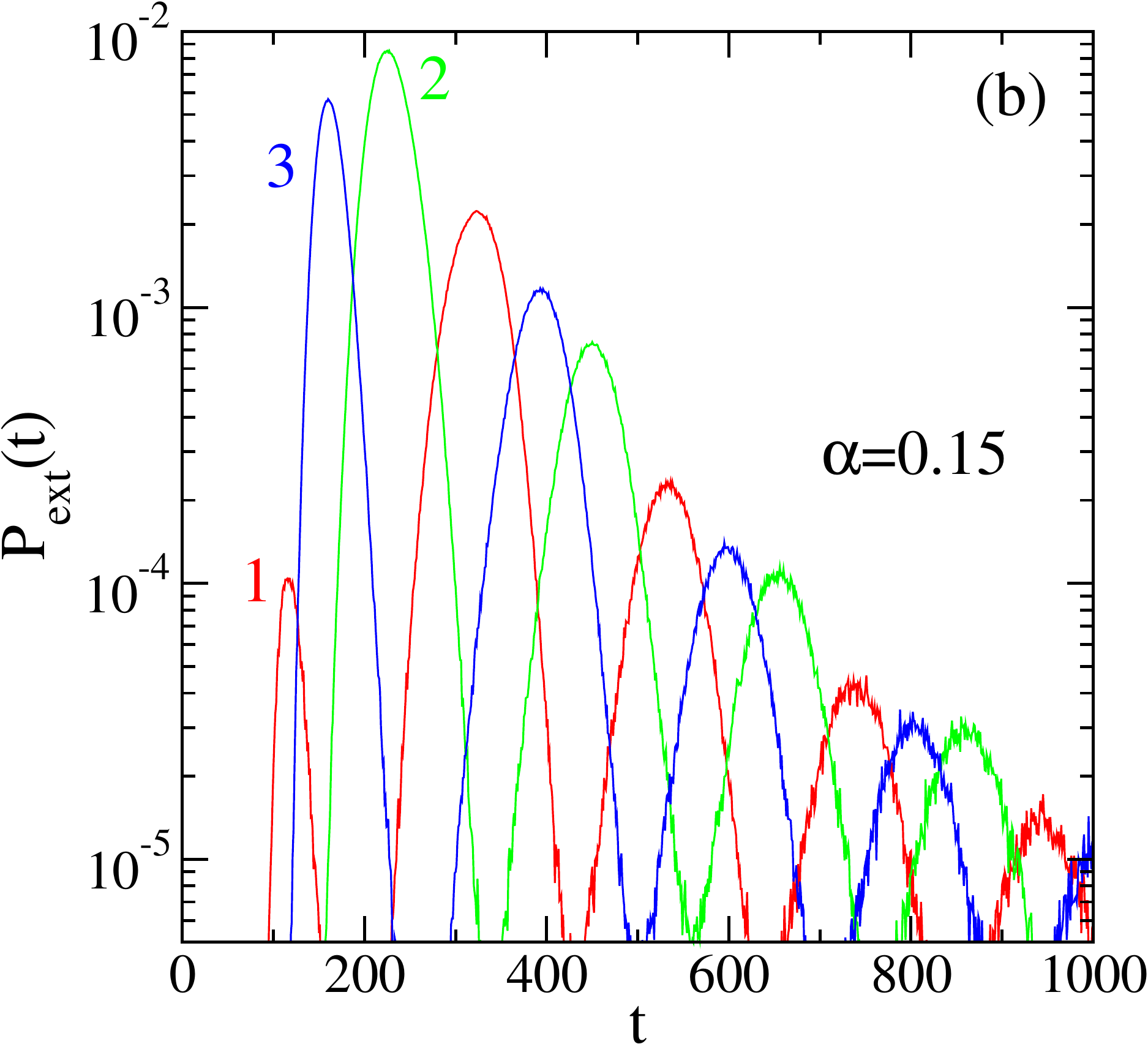}
\caption{(a) Time-dependent extinction probability for each of the three species in systems with $48 \times 48$
sites. The main figure shows the case of habitats where
species 1 has an advantage (indicated by the probability $\alpha = 0.15$ of surviving attacks from its predator species 3), whereas
the inset shows the standard case where all species are treated equally (with $\alpha = 0$). To obtain these probability
distributions 10 million independent runs were performed for every value of $\alpha$. Red: species 1, green: species 2, blue:
species 3. (b) The case $\alpha = 0.15$ in a linear-log plot highlighting the periodicity of the peaks in the probability
distributions.
\label{fig2}
}
\end{figure}

\subsection{Habitat}

For the spatial three-species cyclic game it was shown that spatial heterogeneity in the form of quenched disorder
in the reaction rates does not change qualitatively species coexistence and extinction and has only a minor quantitative
impact on quantities like the typical extinction time \citep{He10,He11}. As we show in the following a spatial heterogeneity
in the form of a structured habitat has a much more profound impact on the properties of systems with cyclic dominance. 

In Figure \ref{fig2}a we compare for a system of $48 \times 48$ sites the time-dependent extinction probabilities for the standard case (inset) as well
as for the case $\alpha = 0.15$ where inside the habitats species 1 has a major advantage that allows its members to
escape an attack unharmed with higher probability. Obviously, for the standard case (the case $\alpha = 0$ in the inset) there is no difference between the
species, and this is reflected by the probability distributions being the same for the three species. For our system of
$48 \times 48$ sites and system parameters that take on the values discussed above 
it takes around 100 time steps after preparation of the system before extinction events show up.
For $\alpha = 0$ the extinction probability increases until $t \approx 300$ when it reaches a maximum. This time of maximal extinction probability
is related to the time the system needs to get organized and form stable space-time patterns. Once these patterns have formed,
the extinction probability decreases and displays an algebraic decay $P_{ext} \sim t^{-\beta}$ with $\beta \approx
2.4$. In presence of habitats the probability distributions are strikingly different, as can be seen in the main panel of Figure \ref{fig2}a.
While species-dependent extinction probabilities are expected, the observation of multiple maxima in these probabilities
is remarkable. These repeated maxima indicate that every species is going periodically through phases where it is susceptible to
die out, followed by phases where it is safe from extinction. These successive maxima, which 
for the parameters used in Figure \ref{fig2} are separated by 206 time steps,
are readily visible in the linear-log plot
shown in Figure \ref{fig2}b. This unexpected behavior goes hand in hand with a higher probability of early extinction. Indeed,
for $\alpha = 0.15$, 99.2\% of all runs see a species going extinct before $t=1 000$, whereas for the standard case 
$\alpha = 0$ species extinction takes place before $t=1000$ in only 83.1\% of the runs.

\begin{figure}
 \centering \includegraphics[width=0.8\columnwidth,clip=true]{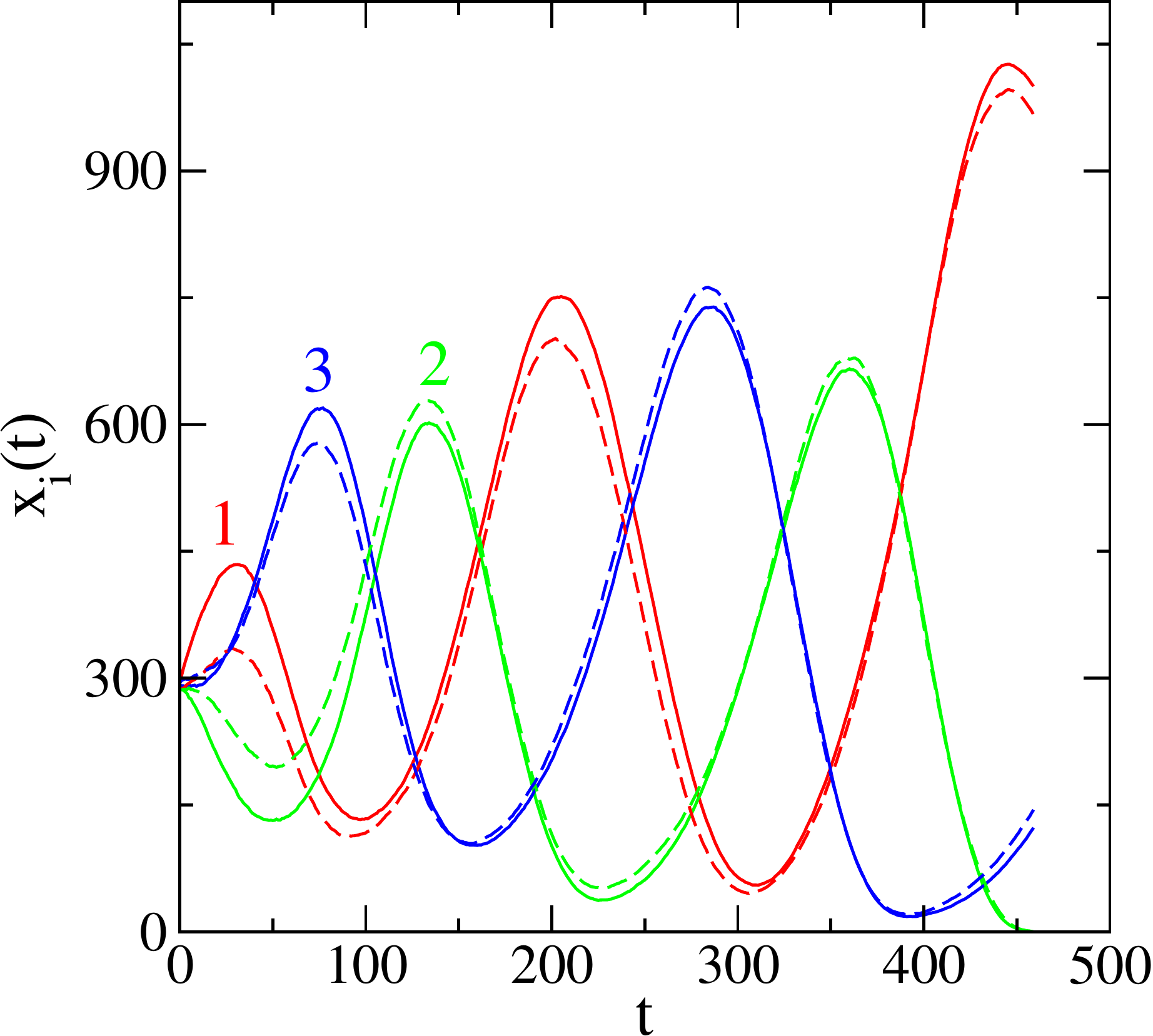}
\caption{Time evolution of the population density $x_i$ of species $i$ ($i=1,2,3$)
both inside the habitats that favor species 1 (full lines) as well outside
these habitats (dashed lines). The number of sites is $48 \times 48$ and $\alpha = 0.15$. These data result from averaging over 100 runs
for which species 2 goes extinct at $t=460$. Red: species 1, green: species 2, blue:
species 3.
\label{fig3}
}
\end{figure}

Periodic oscillations with the same periods as in Figure \ref{fig2} are also showing up in the time-dependent (average) population densities
$x_i(t)$ where $i=1, 2, 3$ labels the three species. This is shown in Figure \ref{fig3} for a subset of runs that are characterized 
by the same extinction event (species 2 goes extinct) at the same time ($t=460$) since preparation of the system. Note that the minima
in the average population density correspond to the maxima in the extinction probability, as it is most likely that noise drives a species
to extinction when the number of individuals of that species is low. Figure \ref{fig3} also reveals differences in the population densities
inside (full lines) and outside (dashed lines) of the habitats that give an advantage to species 1. Whereas one can find more members
of species 1 inside the habitats than outside, it is the opposite for species 2, the prey of predator species 1, while 
the situation is more complicated for species 3 as it depends on
time whether the majority of its members can be found inside or outside of these habitats.

The results discussed so far indicate that a structured habitat has a destabilizing effect on species coexistence 
and creates periodically a situation conductive for species extinction. As discussed above, it is the presence of spiral waves
that enhances the stability of the standard May-Leonard system. In our case the habitats favor locally one of the species and therefore act as sources
of disorder. Consequently, and this has been verified by analyzing snapshots, the emerging space-time patterns in the presence 
of habitats are not given by spiral waves filling the system. Instead, outside the habitats one observes multiple wave fronts that criss cross
that space. Once (part of) a front dominated by species 2 enters a habitat, the individuals from that species are quickly disposed by species 1. 
It is the combination of the wave fronts and the disadvantage species 2 has inside the habitats that yield the oscillations
in the population densities and concomitantly the periodic appearance of peaks in the species extinction probabilities.

The destabilizing effect due to a locally uneven treatment is consistent with the recent observation that in an uneven three-species
May-Leonard model where one species is a less efficient predator than the others the formation of spiral waves is impeded
\citep{Menezes19}. 

We have studied quantitatively the dependence of these features on the linear size of the system $L$ and the probability $\alpha$ that an
individual of species 1 escapes a predator when attacked inside its habitat.
We first note that the destabilizing effect of the habitat is encountered for any value $\alpha > 0$. For a fixed system size
the period of the oscillations (the time between successive peaks of the extinction probability for a given species) shows a slight
dependence on $\alpha$. For example, for $L=48$, the period increase from 200 for $\alpha = 0.08$ to 240 for $\alpha = 0.2$. 
For fixed $\alpha$ a strong increase of the period is observed when increasing the system size.

\begin{figure}
 \centering \includegraphics[width=0.8\columnwidth,clip=true]{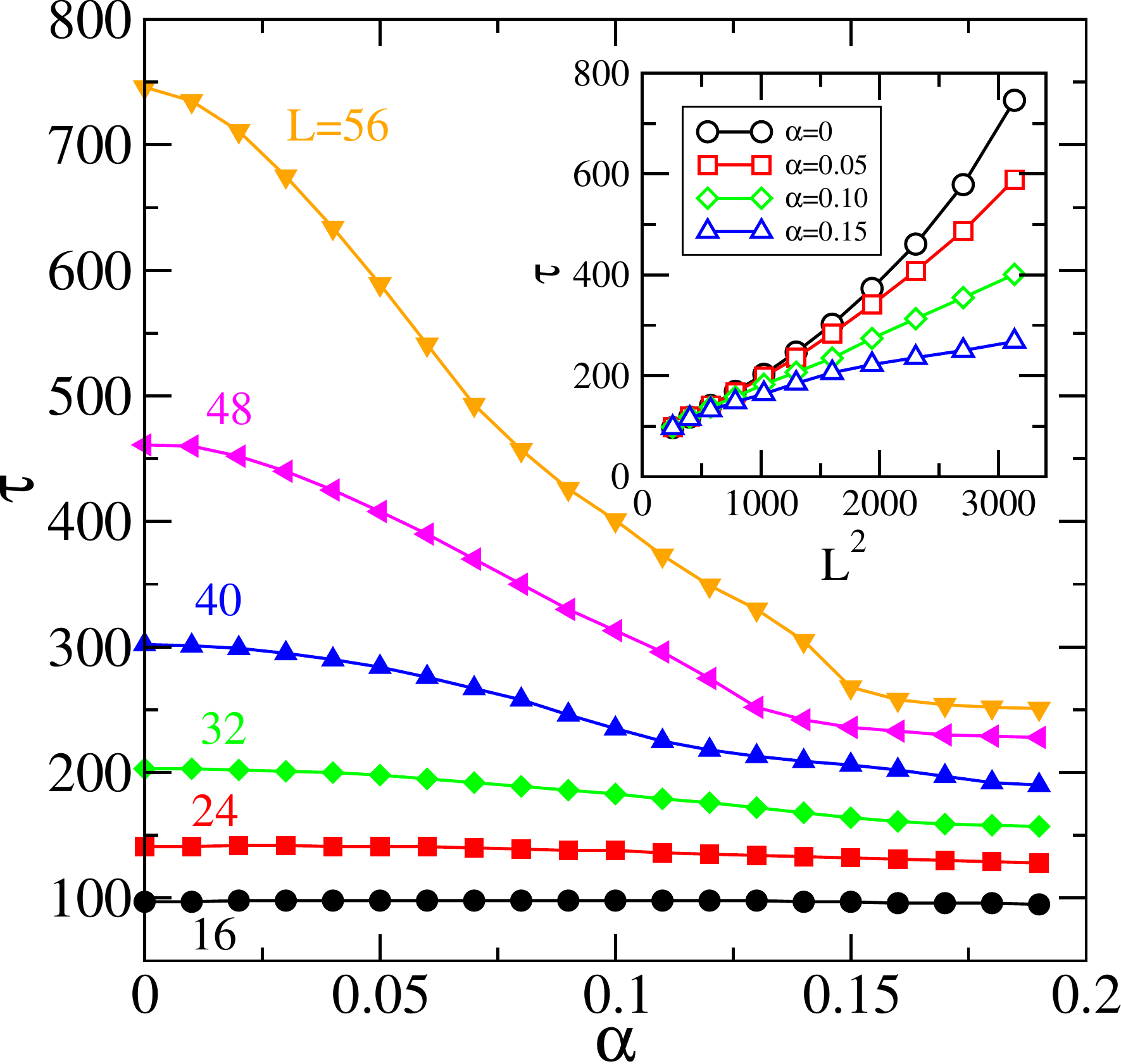}
\caption{
Median extinction time $\tau$ as a function of the habitat modifier $\alpha$ for different linear extents $L$
of the three-species system (ranging from $L=16$ to $L=56$). The inset plots $\tau$ as a function of $L^2$ and shows a transition from stable coexistence
for small values of $\alpha$ to unstable coexistence for larger values of $\alpha$. For every value of $L$ and $\alpha$
one million runs have been performed. Error bars are smaller than the sizes of the symbols.
\label{fig4}
}
\end{figure}

With extinction probability distributions like those shown in Figure \ref{fig2} the mean time to extinction of a species is not
a very good measure as it will be dominated by the extremes. Better suited for our purpose is the median extinction time $\tau$ that we
show in Figure \ref{fig4} as a function of $\alpha$ for various values of $L$ and (in the inset) as a function of $L^2$ for various
values of $\alpha$. Plotting $\tau$ as a function of $\alpha$ reveals for the larger systems different regimes: a first regime for smaller
values of $\alpha$ where $\tau$ displays large changes when $\alpha$ is changed, 
followed by a regime for large $\alpha$ where $\tau$ is largely independent of $\alpha$. Interestingly,
changing the value of $\alpha$ induces a qualitative change of the stability of the system, see inset. Whereas for small $\alpha$ coexistence
is stable ($\tau/L^2$ increases stronger than linear), for $\alpha$ large $\tau/L^2$ increases slower than linear and coexistence is unstable.
These two regimes are separated by a linear relationship between $\tau$ and $L^2$ for $\alpha \approx 0.1$, indicating neutral stability.
Thus a structured environment that locally makes the species uneven has a huge impact on biodiversity of an ecology by
qualitatively changing the stability of coexistence.

\begin{figure}
 \centering \includegraphics[width=0.49\columnwidth,clip=true]{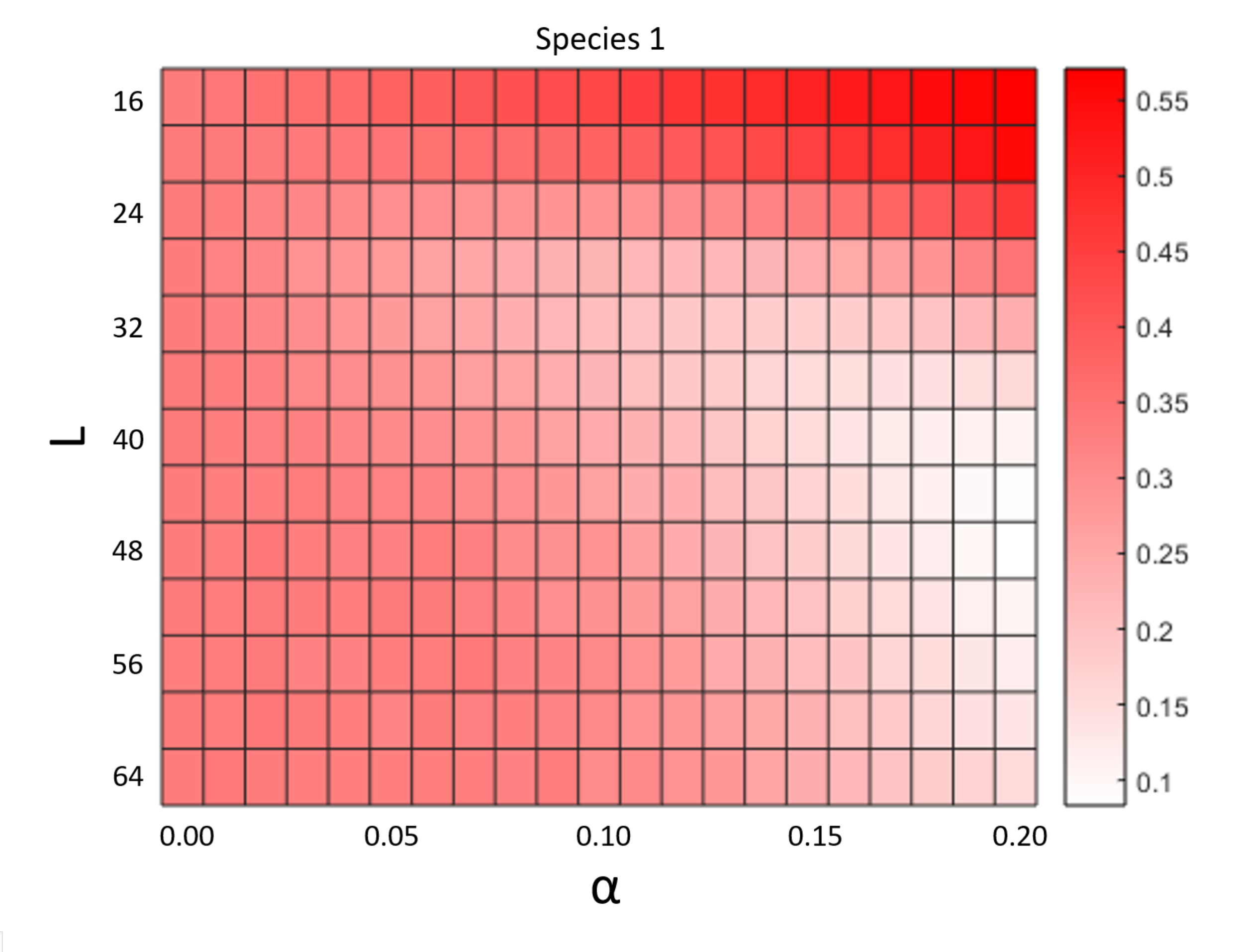}
\includegraphics[width=0.49\columnwidth,clip=true]{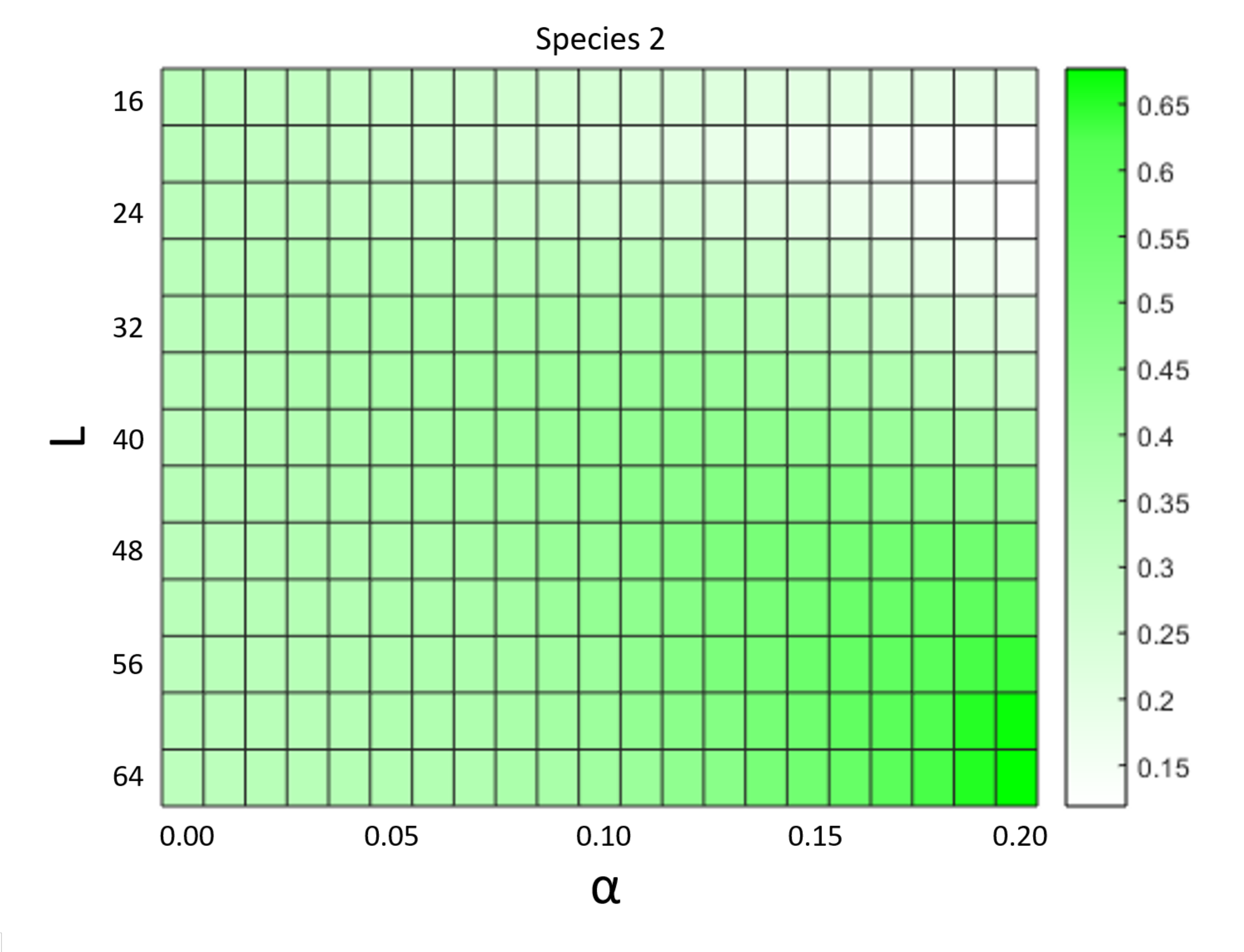}\\
\includegraphics[width=0.49\columnwidth,clip=true]{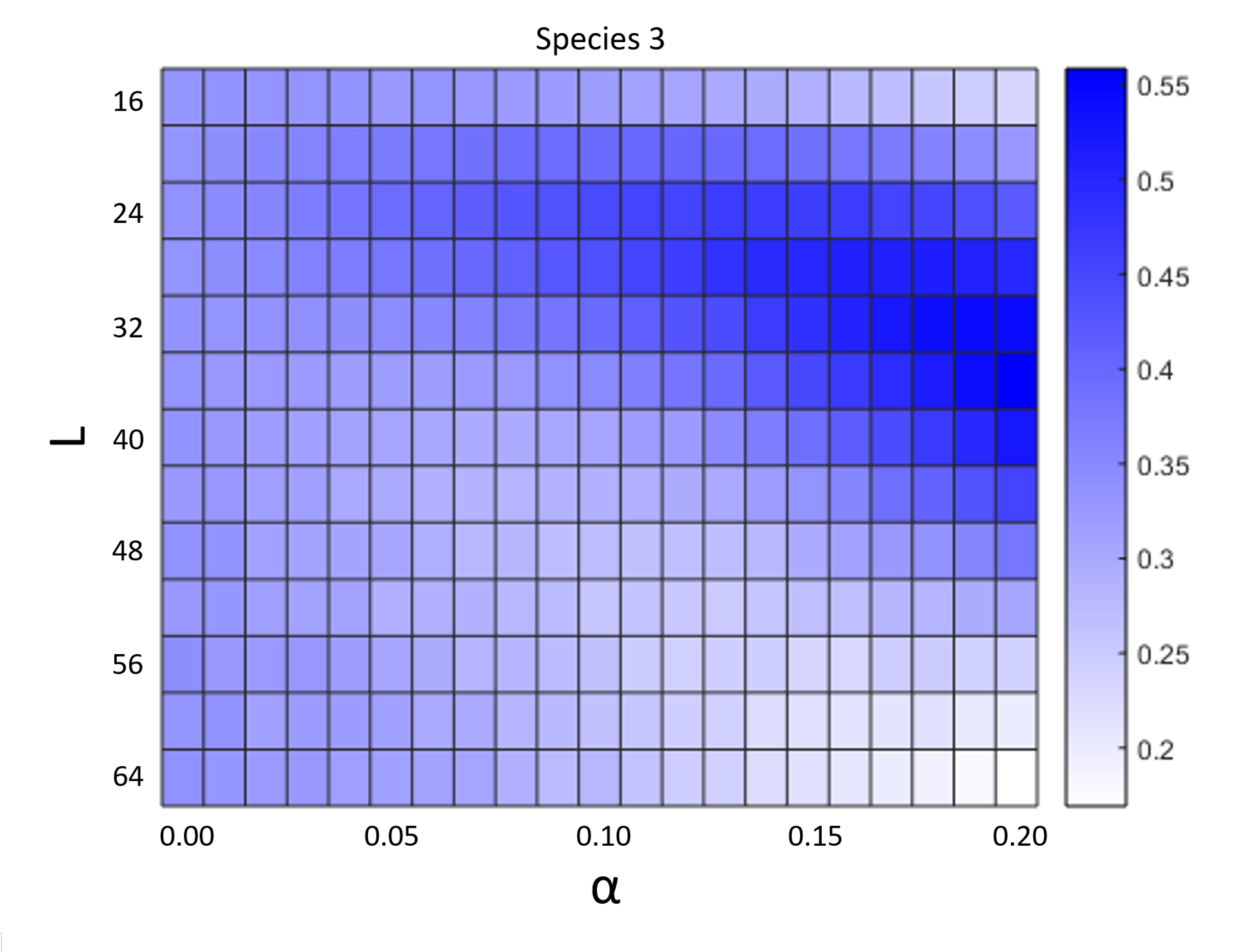}
\caption{Heatmaps showing the probability as a function of $L$ and $\alpha$ that a species dies out first. For large $\alpha$
and large systems $L$ (lower right corner) species 2 has a very large probability to die out first.
\label{fig5}
}
\end{figure}

Finally Figure \ref{fig5} discusses the probability for each species to die out first (the sum of these probabilities being 1). 
For small values of $\alpha$ the existence of the 
habitats does not have a pronounced effect on this probability which is still very much the same for all three species.
For values of $\alpha$ around 0.1 larger effects start to emerge. The advantage provided by habitats makes it very unlikely for
species 1 to die out first for system sizes $L=32$ and larger. This advantage of species 1 is felt strongly by species 2 
as this species is pushed to the brink of extinction by their striving predator. More counterintuitive is
the observation that in smaller systems it is species 3 that has the highest probability to go extinct first, whereas in
the smallest systems species 1 dies out first. This behavior seems to be related to the fact that in systems that are not large
enough we do not see the formation of wave fronts, but it is difficult to understand precisely how this changes the chances
of a species to vanish.

\subsection{Habitat and efficiency}

We have also studied whether individual fitness (efficiency), introduced in a way that it contributes to the probability whether an
attack of a prey by a predator is successful, impacts extinction probabilities. As discussed above, the efficiency of a parent
is inherited by an off-spring, in the sense that the off-spring's efficiency is drawn from a distribution centered around the
parent's efficiency. The fitness parameter not only determines the efficiency as a predator and a prey, it also
allows for evolutionary adaptation.

\begin{figure}
 \centering \includegraphics[width=0.8\columnwidth,clip=true]{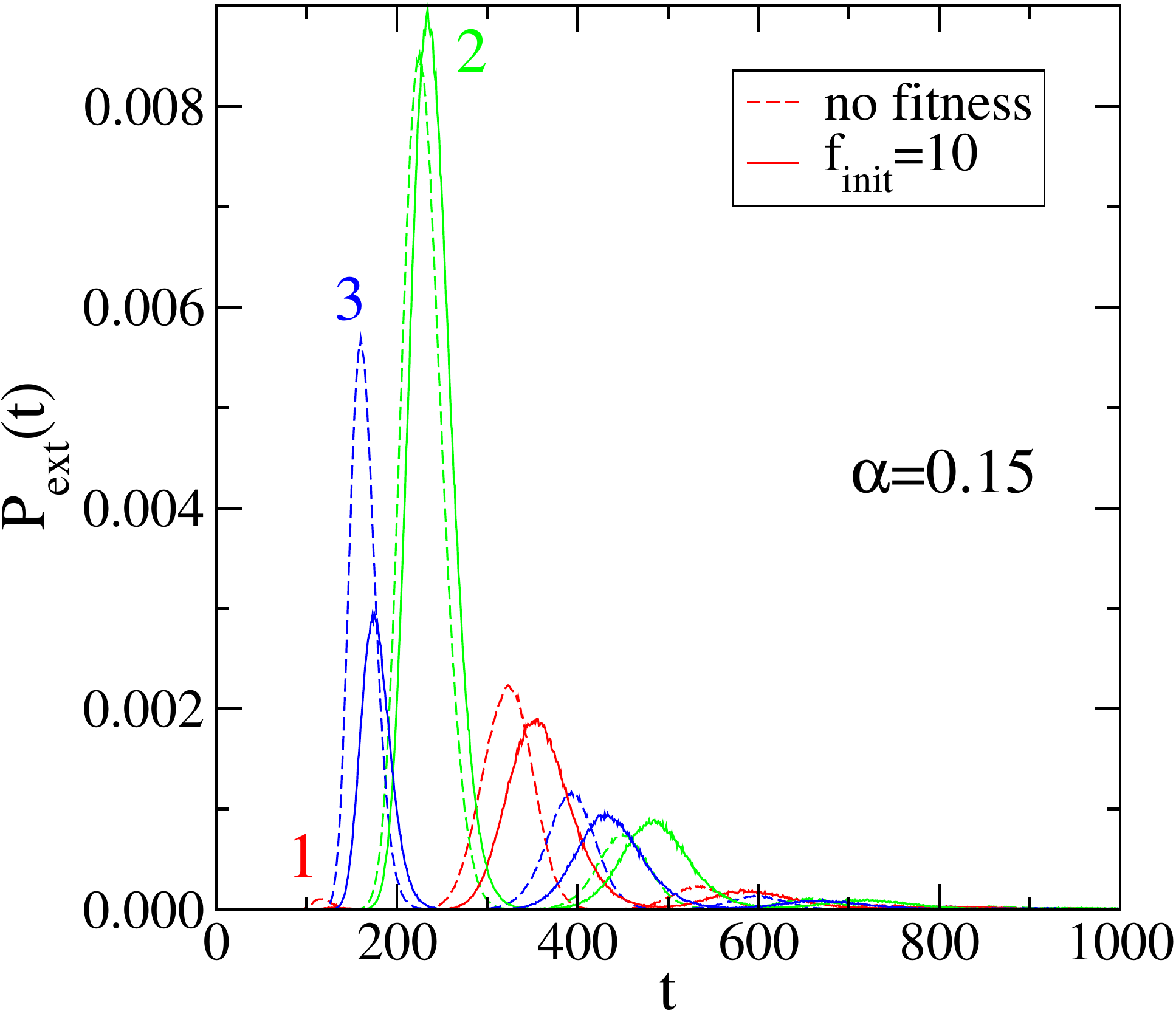}
\caption{Comparison of the time-dependent extinction probabilities without fitness (dashed lines)
and with initial efficiency $f_{init}=10$ (full lines). Other system parameters are $L=48$ and
$\alpha = 0.15$. These probability distributions are obtained after 10 million independent runs.
\label{fig6}
}
\end{figure}

\begin{figure}
 \centering \includegraphics[width=0.8\columnwidth,clip=true]{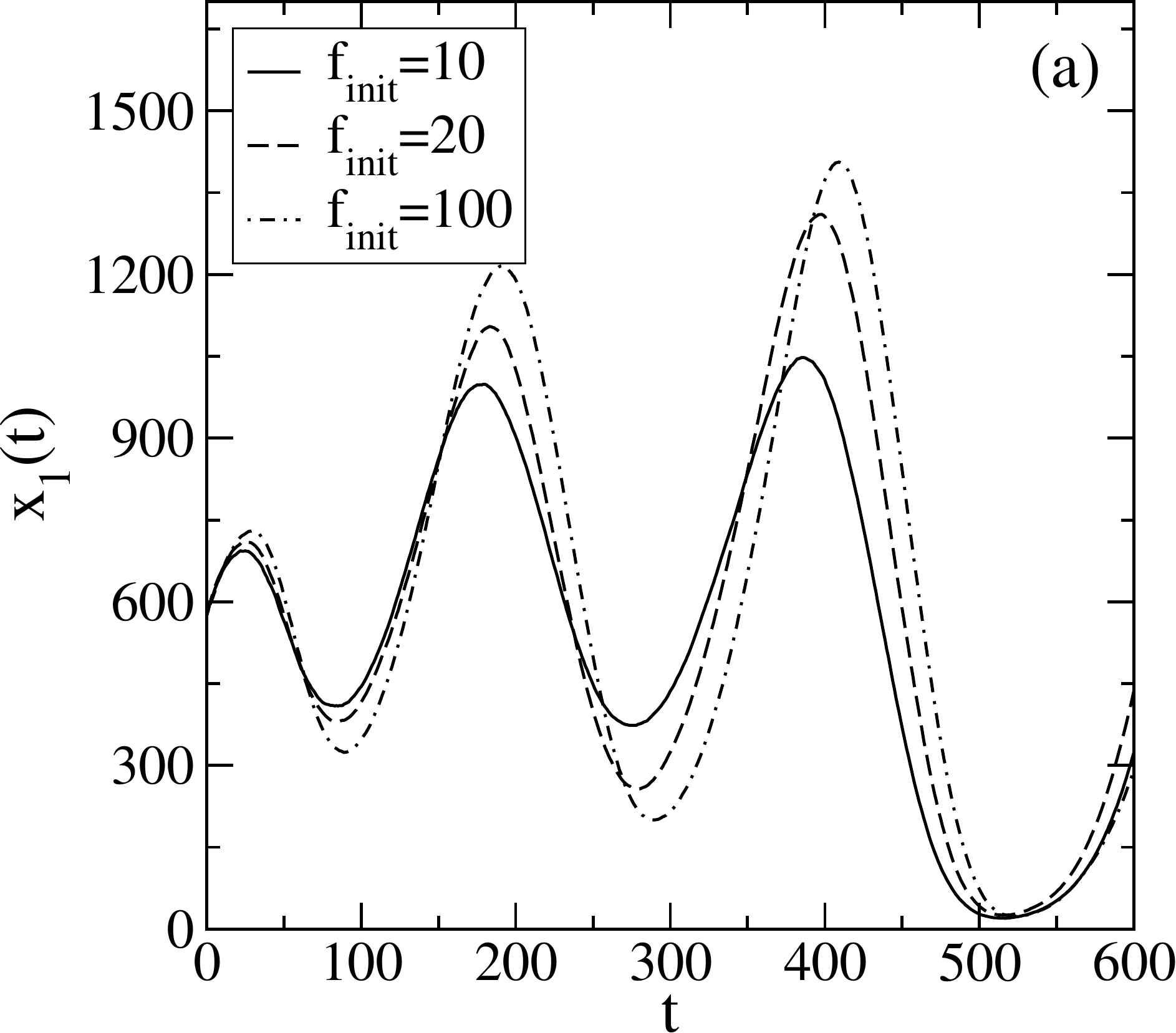}\\[0.2cm]
\hspace*{0.18cm} \includegraphics[width=0.75\columnwidth,clip=true]{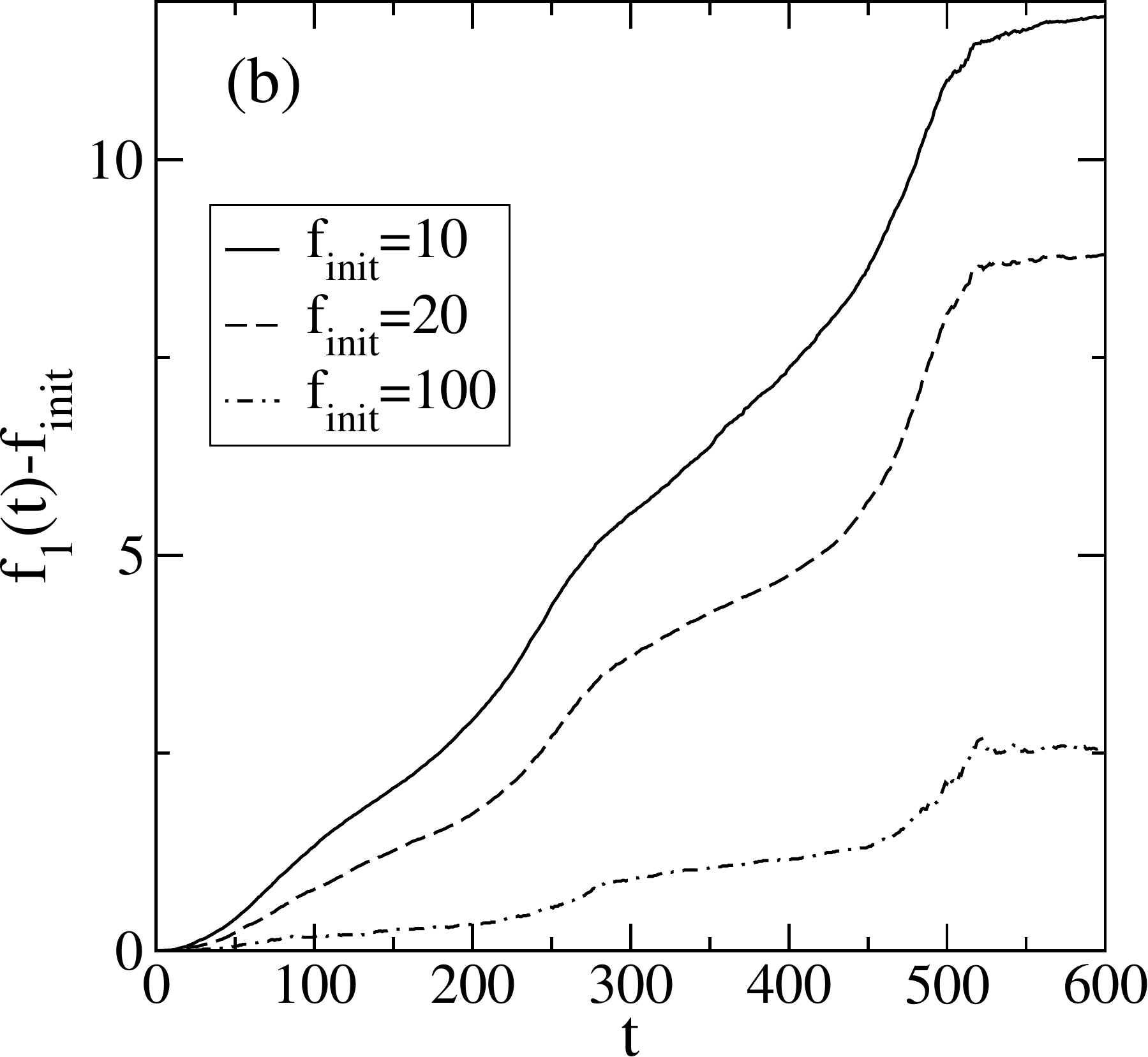}
\caption{(a) Average time-dependent population density and (b) average time-dependent
change in fitness (compared to the initial fitness $f_{init}$) for species 1 obtained
from 100 runs that all end at $t=600$ with the extinction of species 3. Other
system parameters are $L=48$ and $\alpha = 0.15$.
\label{fig7}
}
\end{figure}

As we discuss in Figures \ref{fig6} and \ref{fig7}, including efficiency results in some quantitative changes, but does not modify
the general picture discussed previously, namely that the presence of habitats has a destabilizing effect on temporal patterns
and a negative impact on species coexistence. 

In Figure \ref{fig6} we compare for $\alpha =0.15$ the extinction probabilities without efficiency (dashed lines) with
those where all individuals have been assigned an initial efficiency $f_{init} = 10$. We observe a shift to larger
times for the maxima in the extinction probabilities as well as an increase of the time elapsed between successive
maxima for the same species. These changes can be traced back to the increase over time of the fitness within the different species.

Figure \ref{fig7} shows time-dependent data obtained for different initial efficiencies $f_{init}$. We remind the
reader that the relative change of efficiency between a parent and an off-spring is larger for smaller parent fitness
and that the case without fitness discussed in the previous subsection is recovered in the limit $f_{init}
\longrightarrow \infty$. With that in mind we show in Figure \ref{fig7} the time-dependent population density 
and the time-dependent fitness for species 1 for runs where species 3 dies out after 600 time steps.
Changing the value of $f_{init}$ only yields quantitative changes, but has otherwise a negligible effect
on biodiversity and species extinction. Inspection of the two panels in Figure \ref{fig7} reveals that 
the overall fitness of a species displays pronounced changes at times when the population density decreases.
Whenever a species is under pressure from its predator, the most efficient individuals (in terms of
escaping a predator and catching a prey) persist and produce off-springs that inherit their increased
efficiency.

\begin{figure}
 \centering \includegraphics[width=0.8\columnwidth,clip=true]{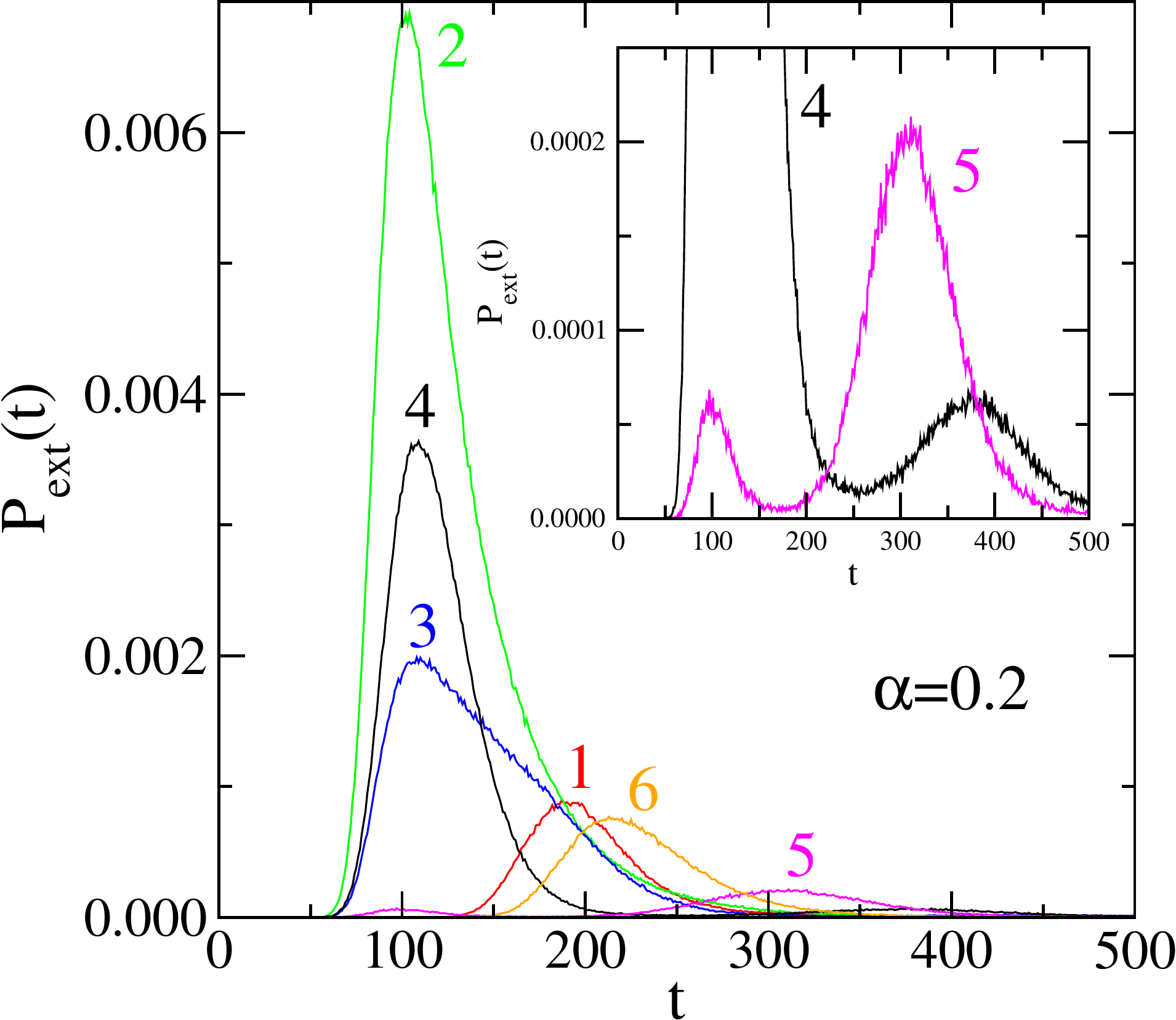}
\caption{Time-dependent extinction probability for each of the species in the six-species system with $48 \times 48$
sites and $\alpha = 0.2$. To obtain these probability
distributions 10 million independent runs were performed. The inset highlights the existence of multiple peaks in
the extinction probabilities of species 4 and 5.
\label{fig8}
}
\end{figure}

\section{The six-species case}
In order to check whether the observed destabilizing effect of a structured habitat is generic we extend our study
to a six-species case that in the spatially homogeneous system also exhibits spiral waves. Whereas the origin of
the spirals in the (6,4) system is the same as for the May-Leonard model (individuals of a given species spontaneously 
arrange in a wave front that follows a front composed of individuals of the one species that is not a prey
of the first one), but there are additional predator-prey relationships between the six species 
that makes this a more involved case.

\begin{figure}
 \centering \includegraphics[width=0.8\columnwidth,clip=true]{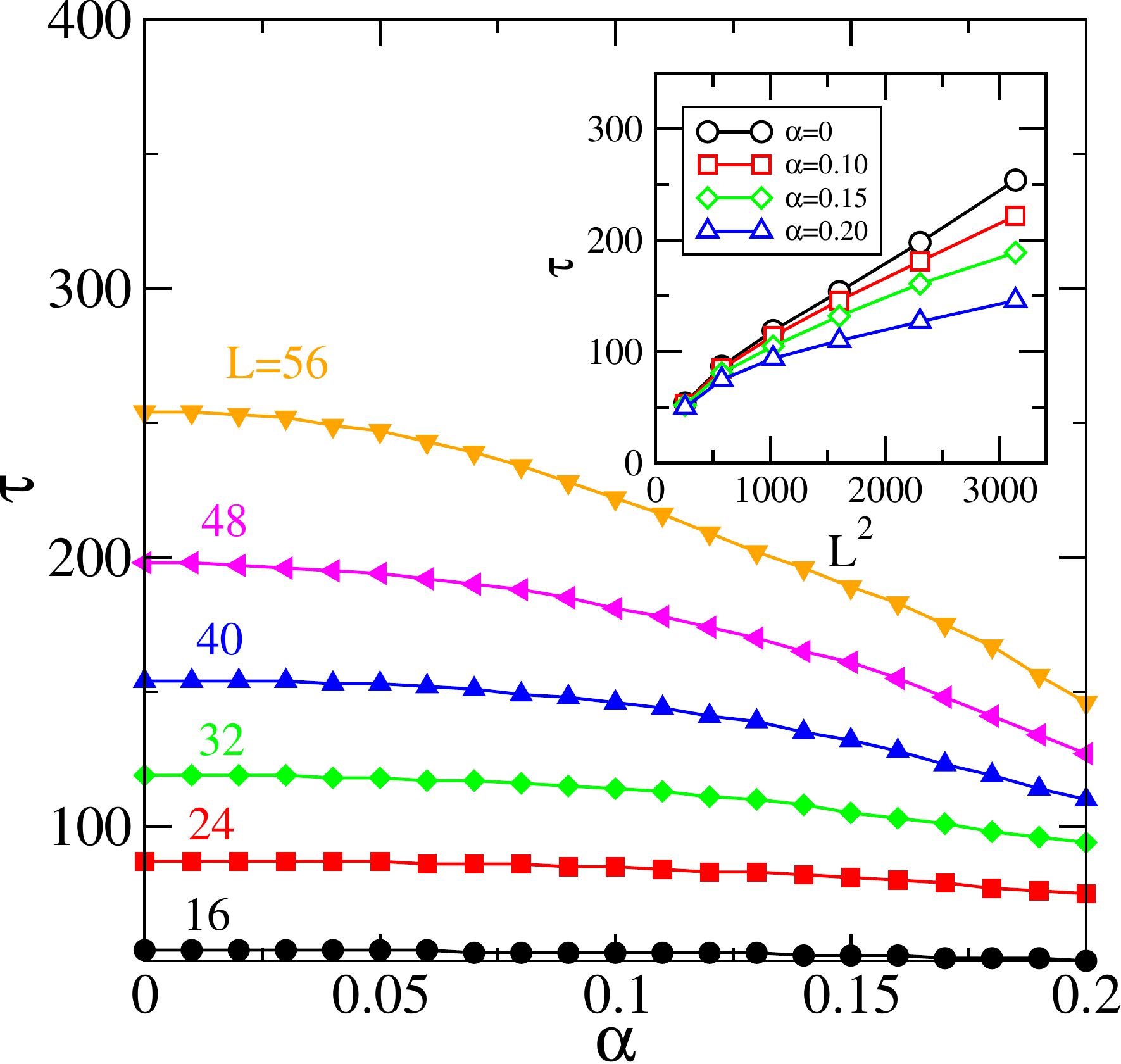}
\caption{
Median extinction time $\tau$ as a function of the habitat modifier $\alpha$ for different linear extents $L$
of the six-species system (ranging from $L=16$ to $L=56$). The inset plots $\tau$ as a function of $L^2$ and shows a transition from stable coexistence
for small values of $\alpha$ to unstable coexistence for larger values of $\alpha$. For every value of $L$ and $\alpha$,
one million runs have been performed. Error bars are smaller than the sizes of the symbols.
\label{fig9}
}
\end{figure}

Inspection of the extinction probabilities as a function of time reveals that in the six-species system with
habitats, that provide an advantage to species 1 through a higher probability to survive unharmed an attack by a predator, the main 
features observed in the three-species case are also observed. This is shown in Figure \ref{fig8} for the case $\alpha = 0.20$ and $L=48$.
As for the three species case the extinction probabilities are different for different species and exhibit a much more complicated
dependence on time as for the homogeneous case $\alpha = 0$ which has again only one broad maximum followed by a power-law decay,
similar to the extinction probabilities for the May-Leonard model shown in the inset of Figure \ref{fig2}a.
The inset in Fig. \ref{fig8} zooms in on the extinction probabilities of
two species and shows the presence of more than one peak in the probability distributions. The data are consistent with maxima that
are separated by constant time intervals, but our computer resources do not allow us to perform a similar quantitative study
of the periodically increasing extinction probabilities as we did for the three-species case.

As before the destabilizing effect of the habitat structure not only yields periodically enhancement of extinction probabilities, it also
changes the median time $\tau$ for an extinction event to occur. Plotting $\tau$ as a function of $\alpha$ for fixed $L$ and as a function
of $L^2$ for fixed $\alpha$, see Figure \ref{fig9}, shows the same trends, albeit not as pronounced, as shown in Figure
\ref{fig4} for the three-species case. From the largest system sizes shown in the inset of Figure \ref{fig9} 
it follows that the median extinction time again reveals
a transition from stable coexistence for $\alpha$ close to zero to unstable coexistence for larger values of $\alpha$.

While we observe the same effects in presence of habitats for three and six species, the effects are less pronounced for the six-species
case. The reason for this can be traced back to the fact that in a homogeneous environment larger systems are needed for the six-species model
to form spiral waves. Still, our data for the six-species case are consistent with those for the three-species case, revealing 
that habitats that locally make the game uneven have generically a debilitating effect on species coexistence.

Finally, we mention that we also introduced efficiency into our six-species system. As for the three-species case only some quantitative
changes (for example minor shifts of peak positions in the extinction probability distributions) are observed.

\section{Summary}
In this paper we addressed the question whether a heterogeneous spatial environment impacts biodiversity and species extinction
in systems characterized by cyclic dominance. Earlier studies for the three-species May-Leonard model focused on quenched spatial
disorder and found that this type of heterogeneity results only in small quantitative changes. Especially it was revealed that quenched
disorder does not have a notable impact on spiral waves which were found to be robust against this type
of perturbations.

In our work we considered a different kind of heterogeneous environment consisting of patches embedded in a matrix. In these patches
one of the species has an advantage (in our implementation that species has a higher probability to escape unharmed an attack from
a predator), whereas in the rest of the system all species are treated equally. Our investigation of two different systems
with cyclic dominance and formation of spiral waves (one being a three-species system, whereas the other is formed by six species)
shows that the structured habitat has a major impact on the observed space-time patterns. Indeed, in this environment well formed spiral waves 
do no fill the system, but instead are replaced by wave fronts criss crossing the space outside the habitats. This behavior is
facilitated by the fact that if a wave front enters one of the habitats, prey of species 1 are quickly removed.
This leads to notable changes to the species extinction probabilities as well as to the median
time to extinction, resulting in a transition between stable coexistence and unstable coexistence.

We also studied how fitness and evolutionary adaptation impacts our systems. For every individual we introduce fitness through a parameter that impinges on
how this individual fares in a predator-prey interaction. In general a species sees a pronounced increase of its average efficiency
in situations where it is under pressure and its population is decreasing. Fitness does however not change dramatically
species coexistence and has only a quantitative effect on extinction times.

Whereas we understand well how a habitat impacts a system with cyclic dominance that form spiral patterns in an homogeneous environment, it is an
open question whether similar effects are observed in other situations characterized by other types of space-time patterns.
We plan to address this question in the future.

\section*{Acknowledgements}
\noindent
{\it Funding:} This work is supported by the US National
Science Foundation through grant DMR-1606814.

\end{document}